\documentclass[aps,twocolumn,superscriptaddress,amsmath,amssymb,prl]{revtex4-1}
\usepackage{graphicx}
\usepackage{dcolumn}
\usepackage{bm}
\usepackage{xcolor}

\usepackage{sidecap}

\usepackage{soul}

\sethlcolor{cyan}

\begin{document}

\title{Controlling bulk conductivity in topological insulators: Key role of anti-site defects}

\author{D.~O.~Scanlon*, P.~D.~C.~King*, R.~P.~Singh, A.~de~la~Torre, S.~McKeown Walker, G.~Balakrishnan, F.~Baumberger, and C.~R.~A.~Catlow\\
\begin{flushleft} [*] Dr.~D.~O.~Scanlon, Prof.~C.~R.~A.~Catlow\\
University College London, Kathleen Lonsdale Materials Chemistry\\
Department of Chemistry, 20 Gordon Street, London WC1H 0AJ, United Kingdom\\
E-mail: d.scanlon@ucl.ac.uk\\
\ \\ \
[*] Dr.~P.~D.~C.~King, A.~de~la~Torre, S.~McKeown Walker, Dr.~F.~Baumberger\\
SUPA, School of Physics and Astronomy, University of St. Andrews\\
 St. Andrews, Fife KY16 9SS, United Kingdom\\
 E-mail: philip.d.c.king@physics.org\\
 \ \\ \
 Dr.~R.~P.~Singh, Dr.~G.~Balakrishnan\\
 Department of Physics, University of Warwick\
 Coventry CV4 7AL, United Kingdom\\
 \ \\ \
 \ \\
 \end{flushleft}}

\begin{abstract}\noindent This is the pre-peer reviewed version of the following article:\\
D.O.~Scanlon, P.D.C.~King, R.P.~Singh, A.~de la Torre, S.~McKeown Walker, G.~Balakrishnan, F.~Baumberger, and C.R.A.~Catlow\\ 
{\it Controlling bulk conductivity in topological insulators: Key role of anti-site defects}\\
Advanced Materials, vol. 24, iss. 16, pp 2154--2158 (2012)\\
which has been published in final form at http://dx.doi.org/10.1002/adma.201200187.
\end{abstract}

%
%
%
%
%
%

\date{\today}

\maketitle

The binary Bi-chalchogenides, Bi$_2$Ch$_3$, are widely regarded as model examples of a recently discovered new form of quantum matter, the three-dimensional topological insulator (TI)~\cite{Zhang:NaturePhys.:5(2009)438--442,Xia:NaturePhys.:5(2009)398--402,Chen:Science:325(2009)178--181,Hsieh:Nature:460(2009)1101--1105}. These compounds host a single spin-helical surface state which is guaranteed to be metallic due to time reversal symmetry, and should be ideal materials with which to realize spintronic and quantum computing applications of TIs~\cite{Hasan:Rev.Mod.Phys.:82(2010)3045--3067}. However, the vast majority of such compounds synthesized to date are not insulators at all, but rather have detrimental metallic bulk conductivity~\cite{Xia:NaturePhys.:5(2009)398--402,Chen:Science:325(2009)178--181}. This is generally accepted to result from unintentional doping by defects, although the nature of the defects responsible across different compounds, as well as strategies to minimize their detrimental role, are surprisingly poorly understood. Here, we present a comprehensive survey of the defect landscape of Bi-chalchogenide TIs from first-principles calculations.  We find that fundamental differences in the energetics of native defect formation in Te- and Se-containing TIs enables precise control of the conductivity across the ternary Bi-Te-Se alloy system. From a systematic angle-resolved photoemission (ARPES)  investigation of such ternary alloys, combined with bulk transport measurements, we demonstrate that this method can be utilized to achieve true topological {\it insulators}, with only a single Dirac cone surface state intersecting the chemical potential. Our microscopic calculations reveal the key role of anti-site defects for achieving this, and predict optimal growth conditions to realize maximally-resistive ternary TIs. 

\begin{figure}
\begin{center}
\includegraphics[width=\columnwidth]{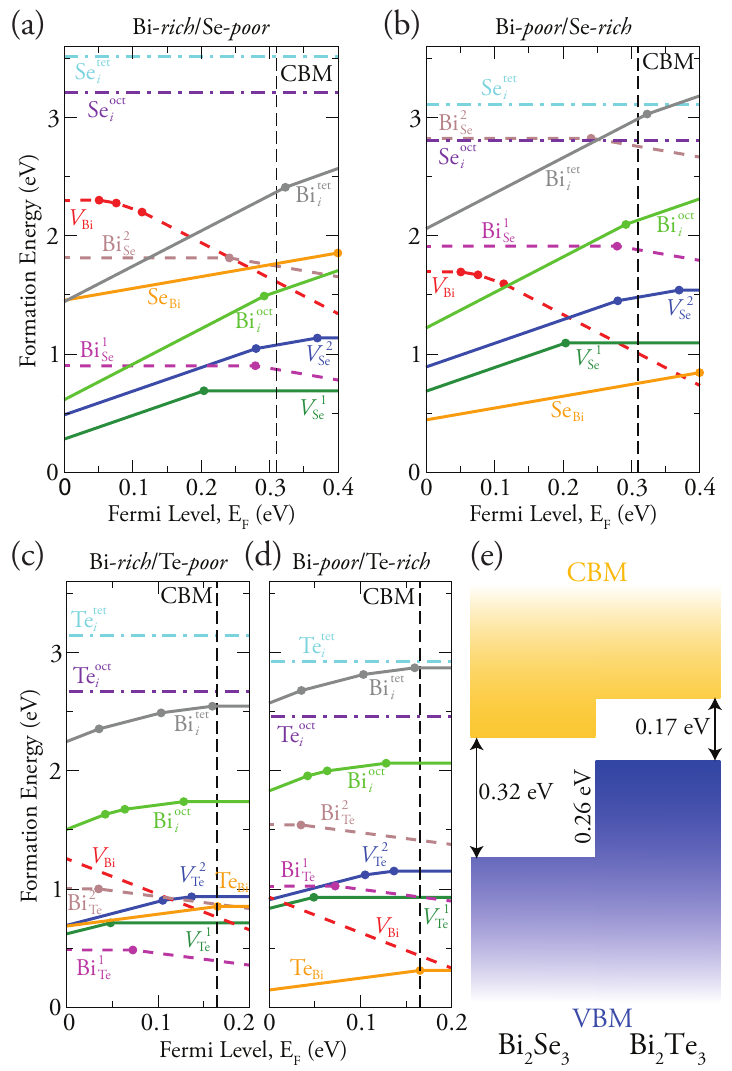}
\caption{ \label{fig:BS_BT_defects} Formation energies, as a function of Fermi level relative to the VBM, of donor (solid lines), acceptor (dashed lines) and electrically-inactive (dot-dashed lines) defects in Bi$_2$Se$_3$ under (a) Bi-rich and (b) Bi-poor conditions. (c,d) Equivalent calculations for Bi$_2$Te$_3$. (e) Calculated valence band offset and resulting band alignment of Bi$_2$Se$_3$ and Bi$_2$Te$_3$.}
\end{center}
\end{figure}
\begin{SCfigure*}
  \centering
\includegraphics[width=1.4\columnwidth]{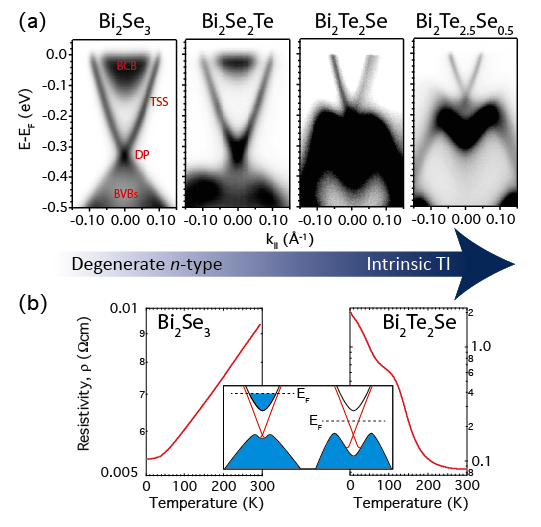}
  \caption{ \label{fig:exp}(a) ARPES measurements of layered Bi-chalchogonides. The bulk valence bands (BVBs) and topological surface state (TSS) are clearly observed for all compounds. Occupied bulk conduction band (BCB) states are only observed for Bi$_2$Se$_3$ and Bi$_2$Se$_2$Te, indicating a transition from degenerately doped semiconductors to true bulk TIs as the Te content is increased. Temperature-dependent resistivity measurements (b) confirm this trend,  showing metallic and insulating behaviour for Bi$_2$Se$_3$ and Bi$_2$Te$_2$Se, respectively. The inset shows a schematic representation of their electronic structure, with the Fermi level located in the conduction band for Bi$_2$Se$_3$ and within the bulk band gap for Bi$_2$Te$_2$Se.}
\end{SCfigure*}

{\bf Figure~\ref{fig:BS_BT_defects}} shows the calculated Fermi-level-dependent formation energies of native defects in the binary TIs Bi$_2$Se$_3$ and Bi$_2$Te$_3$. For Bi$_2$Se$_3$ under Bi-rich/Se-poor conditions (Fig.~\ref{fig:BS_BT_defects}(a)), Se vacancies in the two inequivalent chalcogen layers, $V_{\mathrm{Se}}^1$ and $V_{\mathrm{Se}}^2$, are the dominant donor defects. In particular, $V_{\mathrm{Se}}^1$ has the lowest formation energy of all of the native defects for Fermi levels across the entire bulk band gap, and even up in to the conduction band. This will result in a strong propensity for the formation of $n$-type defects, without significant compensation from $p$-type defects for bulk Fermi levels up to at least 0.1~eV above the conduction band minimum (CBM). This is entirely consistent with our experimental measurements shown in Fig.~\ref{fig:exp}. Our ARPES measurements (Fig.~\ref{fig:exp}(a)), which probe the occupied electronic structure, show not only the bulk valence bands and the topological surface state, but also occupied bulk conduction band states. The Fermi level is located a little over 0.1~eV above the conduction band minimum (CBM), as expected from our calculations, resulting in a large $n$-type conductivity, and a temperature-dependent resistivity (Fig.~\ref{fig:exp}(b)) characteristic of a metal rather than an insulator.

For comparison, we also calculate the formation energies for native defects in Bi$_2$Se$_3$ under Bi-poor/Se-rich conditions (Fig.~\ref{fig:BS_BT_defects}(b)). Compared to the defect energetics under Bi-rich conditions, one would naively expect the formation energy of $V_{\mathrm{Se}}$ to increase, while that of the acceptor-type Bi-vacancy to decrease, and this is indeed seen in our calculations. On this basis alone, much lower residual $n$-type conductivities could be expected under Bi-poor growth conditions, with the Fermi level moving into the bulk band gap (tending towards the intersections of the formation energies of $V_{\mathrm{Se}}^1$ and $V_{\mathrm{Bi}}$). We find, however, that the donor-type Se anti-site defect, Se$_{\mathrm{Bi}}$, becomes the lowest energy defect throughout the band gap. Therefore, as well as the commonly assumed $V_{\mathrm{Se}}$, our calculations indicate that Se$_{\mathrm{Bi}}$ can play a significant role in driving the unintentional conductivity of Bi$_2$Se$_3$. For all possible growth conditions, the lowest energy defect is a donor (either $V_{\mathrm{Se}}^1$ or Se$_{\mathrm{Bi}}$), even when the Fermi level lies at, or slightly above, the CBM. This explains why crystals of Bi$_2$Se$_3$ always display unintentional \textit{n}-type conductivity, which can only be compensated by suitable extrinsic \textit{p}-type doping~\cite{CheckelskyEtAl_PRL2011}. 

The defect physics of Bi$_2$Te$_3$ is rather different, with anti-sites being the dominant defects under both Bi-rich/Te-poor and Bi-poor/Te-rich conditions (Fig.~\ref{fig:BS_BT_defects}(c,d)). Under Bi-poor conditions (Fig.~\ref{fig:BS_BT_defects}(d)), the Te$_{\mathrm{Bi}}$ donor defect has the lowest formation energy of all of the native defects. As for Bi$_2$Se$_3$ under Bi-poor conditions, this will yield unintentional $n$-type conductivity as often observed in experiment~\cite{Chen:Science:325(2009)178--181}, with significant compensation by acceptor  $V_{\mathrm{Bi}}$ centres not expected for Fermi levels within the bulk band gap. However, in contrast to Bi$_2$Se$_3$, under Bi-rich conditions, the formation energy of the acceptor Bi anti-site defect, Bi$_{\mathrm{Te}}^1$, becomes smaller than that of the chalcogen vacancies (Fig.~\ref{fig:BS_BT_defects}(c)). This promotes a natural tendency for unintentional $p$-type conduction when Bi$_2$Te$_3$ is grown under Bi-rich conditions, consistent with both single-crystal growth experiments~\cite{PhysRev.108.1164} as well as recent studies on MBE-grown thin films~\cite{WangEtAl_AM2011}. Thus, our calculations reveal that the defect landscape of Bi$_2$Te$_3$ is dominated by anti-site defects for all growth conditions, and anion vacancies play a much less significant role than in Bi$_2$Se$_3$.

\begin{table}
\begin{center}
\caption{\label{tab:disorder} Anti-site disorder energies, $\Delta E_{\mathrm{ad}} = \frac{1}{2} \left[E_{\mathrm{A}_{\mathrm{B}}} + E_{\mathrm{B}_{\mathrm{A}}} -2E^{\mathrm{pure}}\right]$, where $E_{\mathrm{A}_{\mathrm{B}}}$ is the total energy of a supercell containing an A$_{\mathrm{B}}$ defect, and $E^{\mathrm{pure}}$ is the energy of the stoichiometric Bi$_2$Ch$_2^1$Ch$^2$ supercell.}
\begin{ruledtabular}
\begin{tabular}{lcc}
System          & Species involved  &  $\Delta E_{\mathrm{ad}}$ (eV)  \\ \hline
Bi$_2$Se$_3$    & Bi, Se            &  1.37                           \\ \hline
Bi$_2$Te$_3$    & Bi, Te            &  0.67                           \\\hline
                & Bi, Se            &  1.52                           \\ 
Bi$_2$Te$_2$Se  & Bi, Te            &  0.80                           \\
                & Se, Te            &  0.13                           \\
\end{tabular}
\end{ruledtabular}
\end{center}
\end{table}
In fact, as shown in Table~\ref{tab:disorder}, the energy cost of anti-site disorder in Bi$_2$Te$_3$ is approximately half that of Bi$_2$Se$_3$. This is due to the more similar ionic radii of Bi and Te~\cite{HarmanEtAl_JPCS1957}, and the relatively small differences in electronegativity between the two species: the resulting anti-site defects represent a relatively low-energy configuration. We note that these antsites serve to oppose the ``expected'' polarity of the materials, yielding \textit{p}-type conduction under typical \textit{n}-type growth conditions, and \textit{n}-type samples under typical \textit{p}-type conditions. While growth conditions between these two extremes can yield defect energetics which drive the Fermi level into the bulk band gap, we note that the small size of this energy gap will make achieving robust insulating behaviour very difficult for this compound.

The qualitative differences in the native defect behaviour of Bi$_2$Se$_3$ and Bi$_2$Te$_3$ can be understood from their band alignment. We have computed the natural valence band offsets of these materials using the methodology of Zunger and co workers~\cite{ZhangEtAl_JApplPhys1998, ZhangEtAl-PRL2000}. We find a staggered ``type II''~\cite{Fundamentalsofsemiconductors_1999} offset (Fig.~\ref{fig:BS_BT_defects}(e)), with the valence band maximum (VBM) of Bi$_2$Te$_3$ $0.26\,\mathrm{eV}$ higher in energy than that of Bi$_2$Se$_3$. The smaller ionization potential of Bi$_2$Te$_3$ suggests an increased preference for hole formation~\cite{King:Phys.Rev.B:79(2009)035203}, which fully supports our microscopic calculations. The CBM of Bi$_2$Se$_3$, on the other hand, is only $0.10\,\mathrm{eV}$ below that of Bi$_2$Te$_3$, explaining why both materials display similar \textit{n}-type behaviour under Bi-poor growth conditions. 

Within a conventional semiconductor band engineering methodology, this suggests that alloying Bi$_2$Se$_3$ and Bi$_2$Te$_3$ could be a suitable way to realize bulk insulators, where the topological surface state conduction is no longer shunted by a large residual bulk conductivity. Indeed, transport measurements of the ternary compound Bi$_2$Te$_2$Se have already found a much more insulating bulk resistivity than for the binary compounds~\cite{Ren:Phys.Rev.B:82(2010)241306}, although previous ARPES measurements still showed the occupation of a small number of states at the bottom of the conduction band~\cite{Xu:arXiv:1007.5111:(2010)}. In contrast, our APRES measurements (Fig.~\ref{fig:exp}(a)) show that, upon moving towards Te-rich Bi-Te-Se alloys, the conduction band is readily depleted of carriers. In particular, for both Bi$_2$Te$_2$Se and Bi$_2$Te$_{2.5}$Se$_{0.5}$, only the topological surface state intersects the chemical potential, as desired for a true TI~\cite{BTS_note}.  Both of these compounds exhibit a temperature dependence of their resistivity indicative of bulk insulators. For Bi$_2$Te$_2$Se (Fig.~\ref{fig:exp}(b)), the low-temperature resistivity is as much as two-to-three orders of magnitude higher than in Bi$_2$Se$_3$. However, Bi$_2$Te$_{2.5}$Se$_{0.5}$ is approximately a factor of 6 less resistive than Bi$_2$Te$_2$Se, due to its smaller band gap. 

\begin{figure*}
\begin{center}
\includegraphics[width=\textwidth]{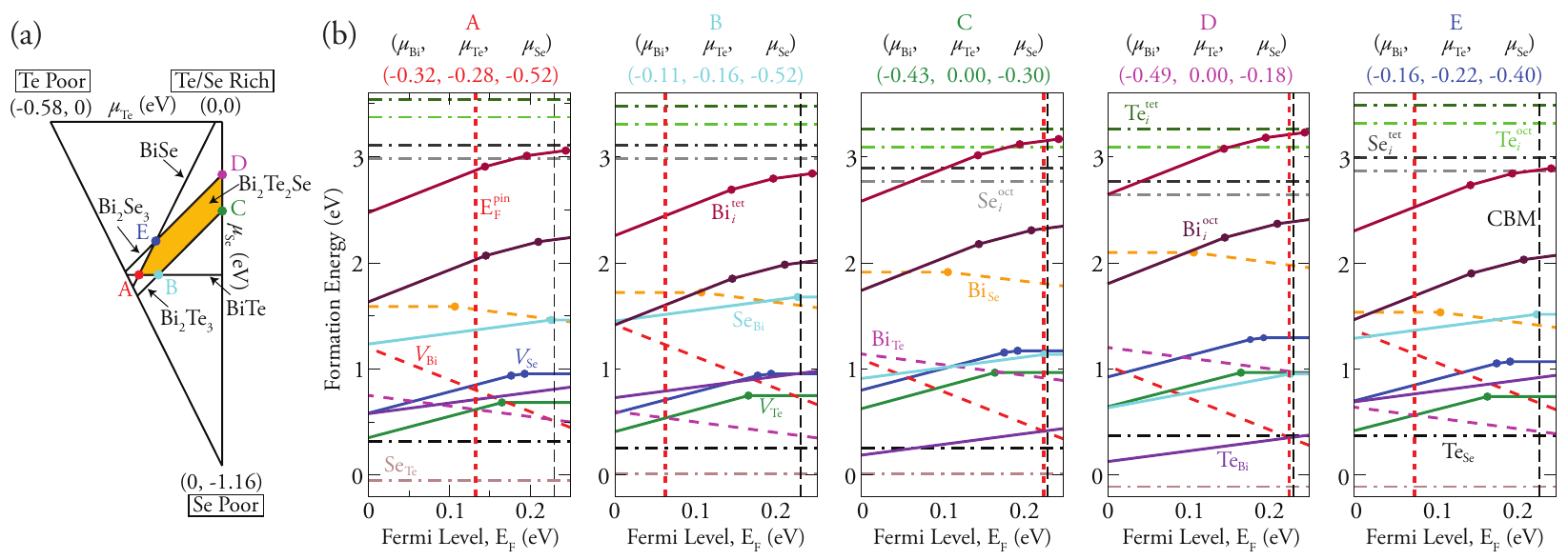}
\caption{ \label{fig:BTS_defects} (a) Calculated phase diagram of Bi$_2$Te$_2$Se as a function of Te and Se chemical potential. (b) Formation energies for defects in Bi$_2$Te$_2$Se, computed for the five sets of chemical potentials shown in (a). }
\end{center}
\end{figure*}

Thus Bi$_2$Te$_2$Se can be seen as a more ideal TI, and we perform explicit calculations for this compound in order to elucidate the microscopic origin of its enhanced resistivity as compared to the binary compounds. Both the Te and Se chemical potentials can be simultaneously varied, subject to the constraints that Bi$_2$Te$_2$Se has lower formation enthalpy than binary compounds of Bi and Te/Se, or than elemental Bi, Te, or Se. Considering these limits, we follow the approach of Walsh \textit{et al.}~\cite{WalshEtAl_JPhysChemC2008} and Persson \textit{et al.}~\cite{PerssonEtAl_PhysRevB2005} to calculate a phase diagram for growth of Bi$_2$Te$_2$Se, shown in Fig.~\ref{fig:BTS_defects}(a). We consider five representative environments in which to calculate formation energies of native defects, shown in Fig.~\ref{fig:BTS_defects}(b).

In addition to the defects considered for the binary compounds, it is now possible to have anion-on-anion anti-site defects (Se$_{\mathrm{Te}}$ and Te$_{\mathrm{Se}}$). In fact, we find that these centres have the lowest formation energy of all native defects across the entire phase diagram. While they have no transition levels within the band gap, and so are electrically inactive, the energy barrier for anion-on-anion anti-site disorder is as low as only $0.13\,\mathrm{eV}$ (Table~\ref{tab:disorder}). It therefore seems inevitable that there will be a significant level of anti-site disorder in these compounds, and Bi$_2$Te$_2$Se alloys will not form as ordered structures as is commonly assumed.

Depending on the growth environment, the lowest energy \textit{p}-type defects are the Bi vacancy or the Bi$_{\mathrm{Te}}$ anti-site, with the dominant \textit{n}-type defects being the Te vacancy or the Te$_{\mathrm{Bi}}$ anti-site. Se vacancies and anion-on-cation anti-sites invariably have higher formation energy than their Te counterparts. In all cases, the formation energy of the dominant donor and acceptor defects cross over within the band gap. Under equilibrium conditions, the Fermi level will tend to be pinned close to this crossing point, represented by the vertical red dotted lines in Fig.~\ref{fig:BTS_defects}(b). For Te/Se-rich conditions (C and D in Fig.~\ref{fig:BTS_defects}(b)), this is very close to the CBM, and so the bulk conductivity will likely still be rather high. However, for Te- and Se-poor growth conditions (A, B, and E in Fig.~\ref{fig:BTS_defects}(b)), this level lies close to the middle of the band gap, where the formation energy for the doubly-charged donor $V_{\mathrm{Te}}$ crosses that of the singly-charged acceptor Bi$_{\mathrm{Te}}$. Our calculations therefore indicate that these (particularly point A) represent the ideal set of growth conditions in which to realize maximally-resistive Bi$_2$Te$_2$Se. Only considering vacancies, the effective Fermi level pinning would be shifted towards much more $n$-type conditions, and so it is again clear that anti-site defects play a key role controlling the unintentional bulk conductivity of ternary, as well as binary, topological insulators. 

\section{Experimental Section}
\textit{First principles calculations:} Density Functional Theory (DFT) calculations were performed using the projector augmented wave method~\cite{KresseAndJoubert_PhysRevB1999} implemented within the VASP code~\cite{KresseAndFurthmuller_PhysRevB1996}. Exchange and correlation were treated within the PBE functional~\cite{PerdewEtAl_PhysRevLett1996}, using a planewave cutoff of $300\,\mathrm{eV}$ and a \textit{k}-point sampling of $10\!\times\!10\!\times\!10$ for the 5 atom tetradymite unit cell. The structure was deemed to be converged when the forces on all of the atoms were less than $0.01\,\mathrm{eV}\,$\AA$^{-1}$. Defect calculations were performed using the method described in Ref.~\cite{Burbano_JACS}, using $4\!\times\!4\!\times\!1$ expansions of the hexagonal representation of the unit cell ({\it i.e.}, $240$ atom supercells) with a 2$\times$2$\times$1 Monkhorst-Pack special \textit{k}-point grid. All calculations include spin-orbit coupling. 

\textit{Experimental details:} Bi$_2$Se$_3$, Bi$_2$Se$_2$Te, Bi$_2$Te$_2$Se, and Bi$_2$Te$_{2.5}$Se$_{0.5}$ crystals were prepared by melting high purity elements (5N) of Bi, Se, and Te in the ratios 2:3:0, 2:2:1, 2:1.05:1.95, and 2:0.5:2.5, respectively. ARPES measurements were performed using Scienta R4000 hemispherical analysers at beamline 5-4 of the Stanford Synchrotron Radiation Lightsource (SSRL) and the CASSIOPEE beamline of synchrotron SOLEIL. The photon energies were between 14 and 18~eV, and the sample temperature was $\sim\!10$~K. Samples were cleaved at the measurement temperature in a pressure better than $3\times10^{-11}$~mbar. Further theoretical and experimental details are given in supplementary information. 

\section{Acknowledgements}
D. O. S. is grateful to the Ramsay Memorial Trust and University College London for the provision of a Ramsay Fellowship. All calculations were made possibe by the UK's HPC Materials Chemistry Consortium, which is funded by the EPSRC (grant no.~EP/F067496). The experimental work was supported by the ERC, Scottish Funding Council, and the EPSRC. SSRL is supported by the US Department of Energy, Office of Basic Energy Sciences. We also acknowledge SOLEIL for provision of synchrotron radiation facilities and we would like to thank Patrick Le Fvre and Amina Taleb-Ibrahimi for assistance in using beamline CASSIOPEE.

\section{Supplementary Information} 

\section{First principles calculations}

\begin{figure*}
\begin{center}
\includegraphics[width=1.5\columnwidth]{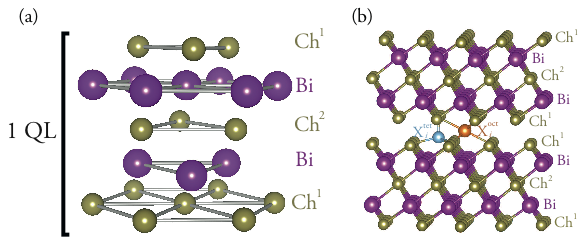}
\caption{ \label{fig:Bi2Ch3_struct} (a) Quintuple-layer (QL) building block of the Bi$_2$Ch$^1_2$Ch$^2$ (Ch = Se,Te) structure. (b) The tetrahedral and octahedral interstitial sites between the quintuple layers.}
\end{center}
\end{figure*}

Prior to performing supercell calculations of the defect formation energies, we calculated the bulk electronic structure of stoichiometric Bi$_2$Se$_3$, Bi$_2$Te$_3$, and Bi$_2$Te$_2$Se using the paramaters as indicated in the manuscript. These calculations yielded indirect band gaps of 0.29~eV and 0.27~eV for Bi$_2$Se$_3$ and Bi$_2$Te$_2$Se, respectively, consistent with experiment. For Bi$_2$Te$_3$, our calculated indirect band gap of 0.08~eV is approximately half of the experimental band gap~\cite{KioupakisEtAl_PRB2010, KimEtAl_PRB2005, Austin_PPS1958}. We corrected this error for the formation energy calculations using a scissors operator. For Bi$_2$Te$_2$Se, we assume an ordered alloy in the calculations. Namely, starting from Bi$_2$Te$_3$, this compound is realized by replacing the middle chalcogen layer of the quintuple-layer structure (Ch$^2$ in Fig.~\ref{fig:Bi2Ch3_struct}(a)) with Se.

We calculate defect formation energies using the method described elsewhere~\cite{Burbano_JACS}, incorporating corrections for the finite size of the supercell~\cite{FreysoldtEtAl_PRL2009} as well as for band-filling resulting from shallow donors~\cite{LanyandZunger_PRB2008}. We have considered the formation of all isolated native defects (see Fig.~\ref{fig:Bi2Ch3_struct}(b) for site definitions): bismuth vacancies ($V_{\mathrm{Bi}}$), both possible anion vacancies ($V_{\mathrm{Ch}}^1$ and $V_{\mathrm{Ch}}^2$), cation and anion interstitials in the tetrahedral and octahedral configurations (Bi/Ch$_i^{\mathrm{tet}}$ and Bi/Ch$_i^{\mathrm{oct}}$),  both Bi on chalcogen antisites (Bi$_{\mathrm{Ch}}^1$ and Bi$_{\mathrm{Ch}}^2$), and the Ch on Bi antisite (Ch$_{\mathrm{Bi}}$). For each compound, we perform our calculations under the limit of Bi-poor and Bi-rich conditions, respectively. Physically, this corresponds to varying the partial pressures during growth, which can be achieved by adjusting the relative Bi and Ch flux as well as the substrate temperature during molecular-beam epitaxy (MBE), or by adjusting the Bi/Ch ratio in the starting mixture for bulk crystal growth. For Bi-rich/Ch-poor conditions, we found the bounding chemical potential values to be set by BiCh (rather than Bi$_2$Ch$_3$) formation.

\section{Sample growth details}
The samples were synthesized starting from powder mixtures in the ratios given in the main manuscript. The mixtures were reacted in sealed, evacuated quartz tubes at 850$^\circ$C for 2 days. For Bi$_2$Se$_3$, this was followed by cooling at 2-3$^\circ$C/h to 650$^\circ$C and annealing at this temperature for 7 days before quenching to room temperature. The Bi$_2$Te$_2$Se, Bi$_2$Te$_{2.5}$Se$_{0.5}$, and Bi$_2$Se$_2$Te   were cooled at 2$^\circ$C/h to 450$^\circ$C, followed by cooling at 50$^\circ$C/h to room temperature. The Bi$_2$Te$_2$Se sample was then resealed in a quartz tube and annealed at 600$^\circ$C for 2 weeks to obtain the insulating behaviour shown in Fig~2 of the main manuscript.

\end{document}